\documentclass[proof]{WileyASNA-v1}

\articletype{ORIGINAL ARTICLE}%

\received{January 11, 2024}
\revised{March 18, 2024}
\accepted{March 19, 2024}

\begin{document}

\title{Gaia23ckh: Symbiotic outburst of the assumed Mira variable V390~Sco}

\author[1]{Jaroslav Merc}

\author[2]{Peter Velez}

\author[3]{Stéphane Charbonnel}

\author[3]{Olivier Garde}

\author[3]{Pascal Le D\^{u}}

\author[3]{Lionel Mulato}

\author[3]{Thomas Petit}

\author[4]{Jan Skowron}

\authormark{MERC \textsc{et al}}

\address[1]{\orgdiv{Astronomical Institute, Faculty of Mathematics and Physics}, \orgname{Charles University}, \orgaddress{V Hole\v{s}ovi\v{c}k{\'a}ch 2, 180 00 Prague, Czech Republic}}

\address[2]{\orgname{Astronomical Ring for Amateur Spectroscopy Group}}

\address[3]{\orgname{2SPOT Southern Spectroscopic Project Observatory Team}}

\address[4]{\orgname{Astronomical Observatory, University of Warsaw}, \orgaddress{Al. Ujazdowskie 4, 00-478 Warszawa, Poland}}

\corres{Jaroslav Merc, Astronomical Institute, Faculty of Mathematics and Physics, Charles University, V Hole\v{s}ovi\v{c}k{\'a}ch 2, 180 00 Prague, Czech Republic.\\ \email{jaroslav.merc@mff.cuni.cz}}


\abstract{The poorly studied variable star V390 Sco, previously classified as a Mira pulsator, was detected in a brightening event by the ESA \textit{Gaia} satellite in September 2023. This work presents an analysis of available archival multifrequency photometric data of this target, along with our spectroscopic observations. Our findings lead to the conclusion that V390 Sco is a new symbiotic star identified by \textit{Gaia}, currently undergoing a classical symbiotic outburst. Additionally, we uncovered three prior outbursts of this system through archival photometry. The outbursts recur approximately every 2330 -- 2400 days, and we hypothesize the periastron passage in an eccentric orbit may trigger them, similarly to the case of BX Mon, DD Mic, or MWC~560. A detailed investigation into the nature of the donor star suggested that V390 Sco is an S-type symbiotic star, likely hosting a less evolved, semiregularly pulsating giant donor, but not a Mira variable. }


\jnlcitation{\cname{%
\author{Merc, J.}, 
\author{Velez, P.}, 
\author{Charbonnel, S.}, 
\author{Garde, O.}, \author{Le D\^{u}, P.}, \author{Mulato, L.}, \author{Petit, T.}, and \author{Skowron, J.}} (\cyear{2024}), 
\ctitle{Gaia23ckh: Symbiotic outburst of the assumed Mira variable V390~Sco}, \cjournal{Astronomische Nachrichten}, \cvol{}.}


\maketitle


\section{Introduction} \label{sec:intro}

\begin{figure*}[t]
\centerline{\includegraphics[width=2\columnwidth]{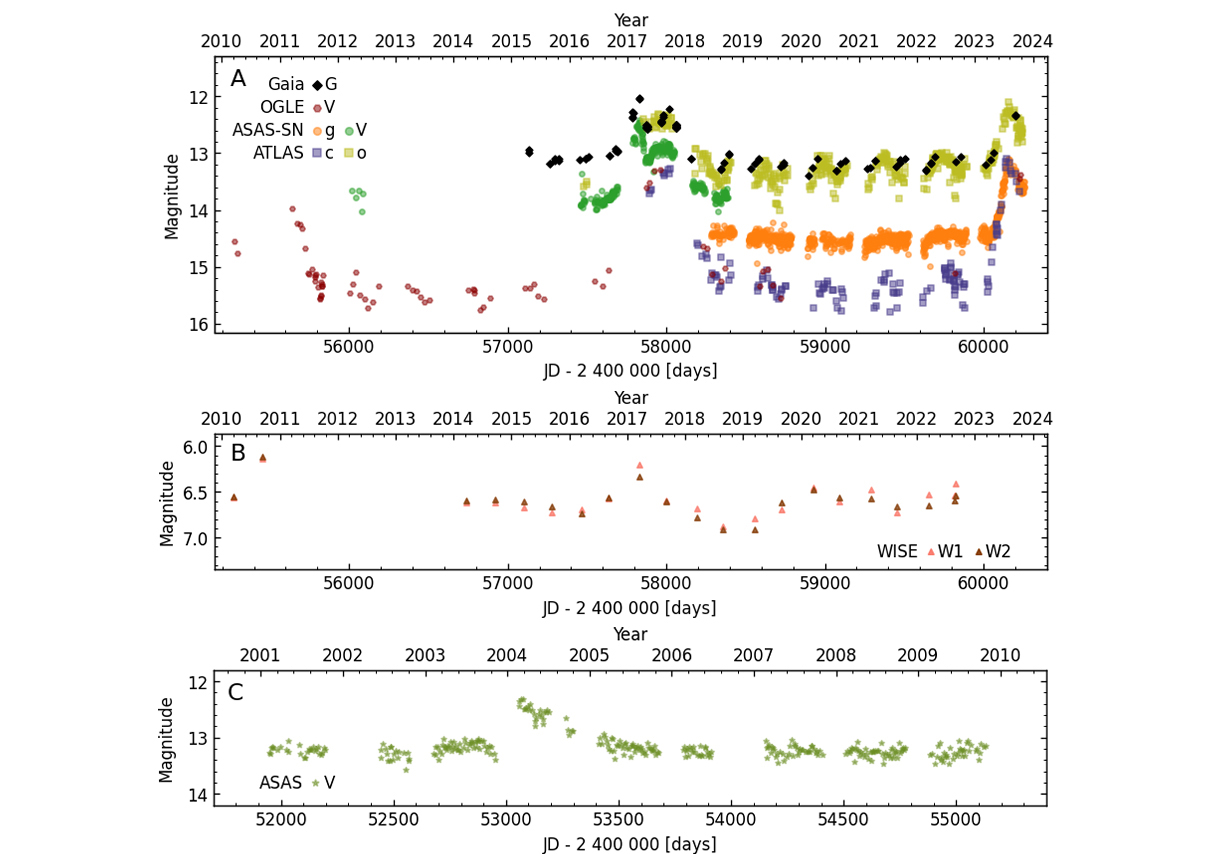}}
\caption{Light curves of V390 Sco. \textbf{A}:~Optical photometry from \textit{Gaia}, ASAS-SN, ATLAS, and OGLE surveys covering three recent outbursts of the target. \textbf{B}:~Infrared light curves from the WISE satellite covering the same timescale as the previous panel. \textbf{C}:~Archival optical photometry from the ASAS survey.\label{fig}}
\end{figure*}

While the European satellite \textit{Gaia} primarily serves as an astrometric mission \citep{2016A&A...595A...1G}, the data it has provided in the data releases \citep[the latest was published in June 2022;][]{2023A&A...674A...1G} have already significantly impacted all fields of observational astronomy. The repetitive observations conducted by \textit{Gaia} enable the mission to also function as an all-sky transient survey, delivering daily reports on changes in the brightness of celestial sources through \textit{Gaia} Science Alerts\footnote{http://gsaweb.ast.cam.ac.uk/alerts/home} \citep[GSA; ][]{2021A&A...652A..76H}.

On September 13, 2023, \textit{Gaia} detected a brightness increase of approximately 0.8 mag in the \textit{G} filter in a source coinciding with a known variable star, V390 Sco. The corresponding GSA was published five days later, labeling the event as Gaia23ckh \citep[=AT 2023swj; $\alpha_{\rm 2000}$ = 17:47:04.90, $\delta_{\rm 2000}$ = -35:59:28.93;][]{2023TNSTR2302....1H}. Only limited information on V390 Sco is available in the published literature. Its long-period variability of at least 2 mag was initially discovered by \citet{1940AnHar..90..207S}, who designated it as HV 6991. Across two sets of observations, they observed a change in the period from 466 days to 420 days. Based on this work, the star was classified as a Mira variable in the General Catalogue of Variable Stars \citep{2017ARep...61...80S} or the SIMBAD database. The only other study that included this object is an analysis of a collection of long-period variables from \textit{Gaia} DR3 \citep{2023A&A...674A..15L}.

In this paper, we examine the available archival photometric data for V390 Sco and provide an analysis of our spectroscopic observations obtained after the publication of the GSA. The paper is organized as follows. Section \ref{sec:photometry} discusses the photometric behavior of the source, while Section \ref{sec:spectra} details the analysis of spectroscopic data and discusses the classification of the object as a symbiotic star. The nature of the donor star in the studied system is explored in Section \ref{sec:giant}, and we conclude by summarizing our results and presenting the conclusions in Section \ref{sec:conclusions}.

\section{Photometric behaviour}\label{sec:photometry}

To investigate the ongoing brightening of V390 Sco and explore its long-term photometric behavior, we complemented the \textit{Gaia} $G$ light curve obtained from the GSA website with photometric data from the All-Sky Automated Survey for Supernovae \citep[ASAS-SN;][]{2014ApJ...788...48S, 2017PASP..129j4502K}, the All-Sky Automated Survey \citep[ASAS;][]{1997AcA....47..467P}, and the Asteroid Terrestrial-impact Last Alert System (ATLAS) project \citep{2018PASP..130f4505T,2020PASP..132h5002S} obtained from the ATLAS Forced Photometry server \citep{2021TNSAN...7....1S}. We also included the $V$ filter data from the Optical Gravitational Lensing Experiment survey \citep[OGLE~IV; ][]{2015AcA....65....1U} into our analysis. The much higher cadence OGLE data in the $I$ filter were unfortunately saturated for this star and were, therefore, not used in this work. Additionally, we analyzed infrared photometry from the Wide-field Infrared Satellite Explorer \citep[WISE;][]{2010AJ....140.1868W}, secured in the \textit{W1} and \textit{W2} filters as part of the reactivated NEOWISE project \citep{2011ApJ...731...53M,2014ApJ...792...30M}. The daily averages are calculated for ASAS, ASAS-SN, and ATLAS data. In the case of WISE data, several exposures obtained in 2-3 day windows are averaged.

\subsection{Outbursting events}

The light curves of the object in various filters are presented in Fig. \ref{fig}. Analysis of observations from the ATLAS and ASAS-SN surveys reveals that the current brightening started in early March 2023 (around JD\,2\,460\,030), though it went unnoticed until the publication of the GSA in September. An earlier detection by \textit{Gaia} was hindered by the scanning law of the satellite, and the brightening was not alerted by other surveys, likely due to the prior classification of a star as a variable. The observations indicate that the maximum brightness was achieved in August 2023 (around JD\,2\,460\,160), before the GSA publication. The amplitude of the current brightening was around 2.3 mag, 1.3 mag, and 1.2 mag in the ATLAS $c$, ASAS-SN $g$, and ATLAS $o$ filters, respectively. The approximate amplitudes in the \textit{Gaia} $G$ and OGLE $V$ were $\sim$0.9 mag and $\sim$1.9 mag, respectively.

\begin{figure}[t]
\centerline{\includegraphics[width=\columnwidth]{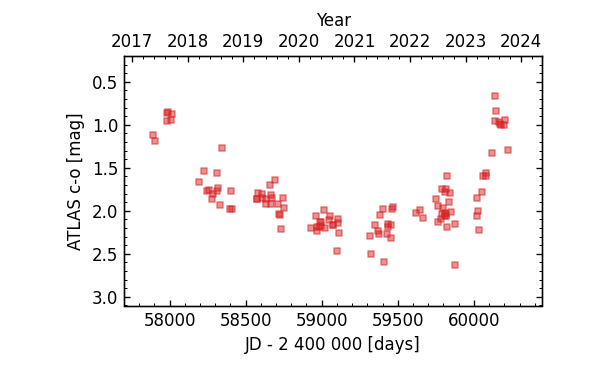}}
\caption{Color evolution of V390 Sco derived from ATLAS data, illustrating the temporal variation of $c$-$o$ colors during the recent two brightenings. This figure utilizes pairs of closest observations in $c$ and $o$ filters, considering only those with a time difference < 5 days. The average time difference between $o$ and $c$ data was 2.2 days. \label{fig:colors}}
\end{figure}

\begin{figure}[t]
\centerline{\includegraphics[width=\columnwidth]{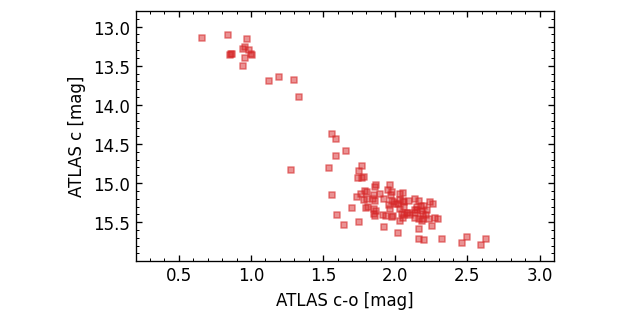}}
\caption{Evolution of V390 Sco in the HR-like diagram. The figure illustrates the correlation between ATLAS $c$-$o$ colors and brightness in ATLAS $c$ filter. The data were selected in the same way as for Fig. \ref{fig:colors}. \label{fig:colors_hr}}
\end{figure}

During the brightening, the object exhibited significantly bluer colors, with ATLAS $c$-$o$ changing from approximately 2.0 mag before the outburst to about 0.9 mag at the outburst maximum (Fig. \ref{fig:colors} and \ref{fig:colors_hr}). The change in colors is also confirmed by the uncalibrated low-resolution BP/RP slitless spectra \citep[R $\sim$ 100; see more in][]{2016A&A...595A...1G,2023A&A...674A...1G} available directly on the GSA website.

In addition to the recent brightening, three other events are evident in the available light curves. A prior event occurred in 2017 and is clearly visible in ASAS-SN, ATLAS, OGLE, and \textit{Gaia} light curves (Fig. \ref{fig}A). It is displaying similar properties (amplitudes and color evolution; see also Fig. \ref{fig:colors}) to the 2023 detection. Interestingly, Gaia and ASAS-SN observations reveal a secondary maximum approximately 150 days after the main peak. The increase and decrease in brightness during this event are not as abrupt as in the main outburst. While such behavior is not observed in the 2023 brightening, it may have occurred during the seasonal gap in the observations. It is worth noting that the 2017 brightening is also distinguishable in the WISE infrared light curves (Fig. \ref{fig}B). Unfortunately, the same data for the recent brightening are not yet available.

Another outburst of V390 Sco was identified in ASAS data in 2004 (Fig. \ref{fig}C). The amplitude in the $V$ filter is approximately 0.9 mag; however, it is important to note that these data might be contaminated by neighboring stars influencing the measured amplitudes of variability. OGLE data obtained in the $V$ filter revealed additional outburst that occurred in 2010. It is difficult to estimate its amplitude from the available data as the maximum falls into the seasonal gap. This brightening is also confirmed by a comparison of recent WISE light curves (covering years 2014 -- 2022) and data from the AllWISE catalog that were obtained in 2010 (Fig. \ref{fig}B).

In general, the described behavior does not resemble the pulsations of a single Mira but rather reflects the characteristics of symbiotic stars — interacting binaries comprising a red giant donor and a white dwarf accretor \citep[see, e.g.,][]{2012BaltA..21....5M,2019arXiv190901389M}. The amplitudes, durations, and recurrence timescales of the outburst events observed in V390~Sco closely resemble the behavior of classical symbiotic stars. We discuss the symbiotic classification in more detail in Sec. \ref{sec:spectra}. 

Interestingly, the time intervals between the four observed outbursts appear to be similar ($\sim$ 2330 -- 2400 days). This suggests a possible connection between these periodic outbursts and the enhanced accretion rate during the periastron passage of the hot component, particularly if the eccentricity of the orbit is non-zero. The recurrence period of the outbursts would then align with the orbital period of the system. However, confirmation of this hypothesis requires spectroscopic monitoring.

Similar behavior has been observed in the cases of BX Mon and DD Mic by \citet{2013AcA....63..405G}, MWC 560 \citep[e.g.,][]{2007A&A...463..703G}, and possibly also in the case of V2905 Sgr \citep{2013AcA....63..405G,2023ATel16257....1M}, V840 Cen, Hen 3-1103 \citep{2013AcA....63..405G}, and FN Sgr \citep{2023A&A...675A.140M}. However, the orbital eccentricity in these latter cases is not yet known.

\subsection{Quiescent variability}

\citet{1940AnHar..90..207S} reported periodic variability with a period ranging from 420 to 466 days in the light curves of V390 Sco. As mentioned earlier, this periodicity led to the classification of the star as a Mira variable. We attempted to analyze the quiescent light curves of the star, i.e., those obtained between the brightenings, to search for any periodic variability. The ASAS and ASAS-SN light curves, unfortunately, revealed no apparent variability. This lack of detection may be attributed to the surveys' poor angular resolution, resulting in issues with photometric contamination from nearby stars, as well as the relative faintness of the target in quiescence.

\begin{figure}[t]
\centerline{\includegraphics[width=\columnwidth]{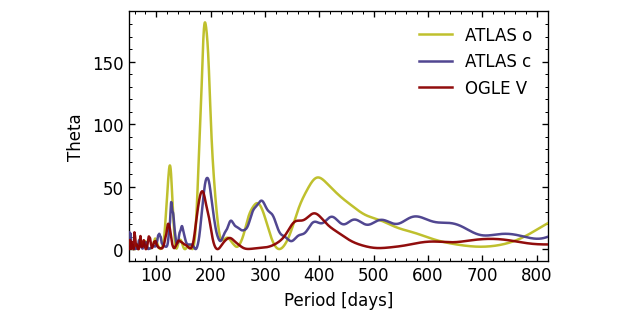}}
\caption{Lomb-Scargle periodogram of ATLAS $o$ and $c$ and OGLE $V$ data. A dominant period is close to 190 days. The period of around 125 days is an alias. Theta value for OGLE $V$ was multiplied by a factor of 5 for clarity. \label{fig:period}}
\end{figure}

\begin{figure}[t]
\centerline{\includegraphics[width=\columnwidth]{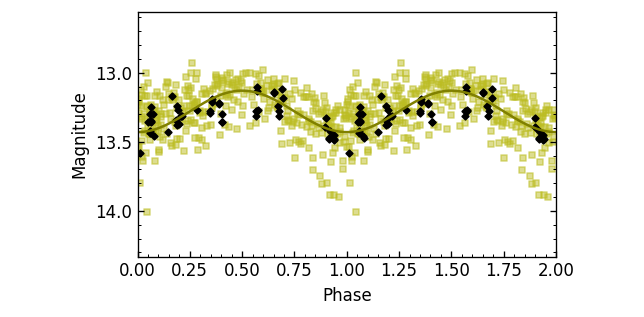}}
\caption{Light curves of V390 Sco in ATLAS $o$ and \textit{Gaia} $G$ filters (shifted by +0.18 mag for clarity) phased with a period of 189.5 days. The colors of symbols are the same as in Fig.~\ref{fig}. The solid line represents a sinusoidal curve with the same period to guide the eye. \label{fig:phased}}
\end{figure}

Conversely, periodic variability is evident in the ATLAS light curves, whose pixel scale is significantly smaller (1.86" vs. 8.00" of ASAS-SN). Period analysis using the Lomb-Scargle method \citep{1976Ap&SS..39..447L,1982ApJ...263..835S} in the Peranso software \citep{2016AN....337..239P} disclosed a dominant period of approximately 190 days (Fig.~\ref{fig:period}). This period is clearly detected in the $o$ filter (189.5$\pm$2.7 days; $\lambda_{\rm eff} = 6630$\,\AA) light curves and is discernible also in the $c$ filter (193.5$\pm$6.9 days; $\lambda_{\rm eff} = 5182$\,\AA), albeit the data quality is lower as the star is fainter in this part of the spectrum. We detected a comparable period of 184.5$\pm$5.4 days in the OGLE observations spanning the period between the 2010 and 2017 outbursts, albeit with a significantly reduced number of data points.. Additionally, the period from the ATLAS light curves aligns with the \textit{Gaia} data. However, due to the scanning law of the satellite, observations of the star between the two outbursts were consistently acquired during brightness increases, complicating the standalone period analysis of \textit{Gaia} data. The light curves in $o$ and $G$ filters, phased with the dominant period derived from the analysis of ATLAS data, are presented in Fig.~\ref{fig:phased}. The time intervals corresponding to the outbursts of the star were omitted in this analysis.

Interestingly, the period we obtained from the analyzed data is approximately half of the period reported earlier. Due to unknown time sampling and the number of observation epochs in the original data, it is challenging to ascertain if this difference is attributed to sampling. It also remains an open question if the variability could be associated with the outburst activity, given that the reported amplitude of 2 mag is much higher than currently observed in quiescence in the available dataset. Nonetheless, the observed variability does not resemble that of Mira variables. We investigate the nature of the giant star in more detail in Sec. \ref{sec:giant}.

We note here that while obtaining reasonable results from the period analysis of the quiescent WISE data is challenging due to data sparsity, there appears to be a potential additional minor maximum around 2020, approximately 1100 days after the maximum observed during the 2017 outburst. This timing aligns closely with half of the period between the outbursts (a possible orbital period of the system).

Such a relationship between the periodicities observed in blue and red bands is often indicative of the giant star either filling or being close to filling its Roche lobe, resulting in the detection of two minima per orbital period due to the ellipsoidal effect. However, under reasonable assumptions about the masses of the components of the system, assuming an orbital period of 2\,350 days would imply an implausibly large radius for the giant star (a few $\times$ 10$^2$\,R$\odot$), which contradicts other findings from this work.

Alternatively, it is possible that the infrared flux was lower after the 2017 outburst (during the period from 2018 to 2020) than during 'normal' quiescence, potentially creating a deceptive appearance of an additional maximum. Only long-term spectroscopic monitoring, ideally in the infrared, can help further refine the orbital period of the system.

\section{Spectra and symbiotic classification}\label{sec:spectra}

\begin{figure}[t]
\centerline{\includegraphics[width=\columnwidth]{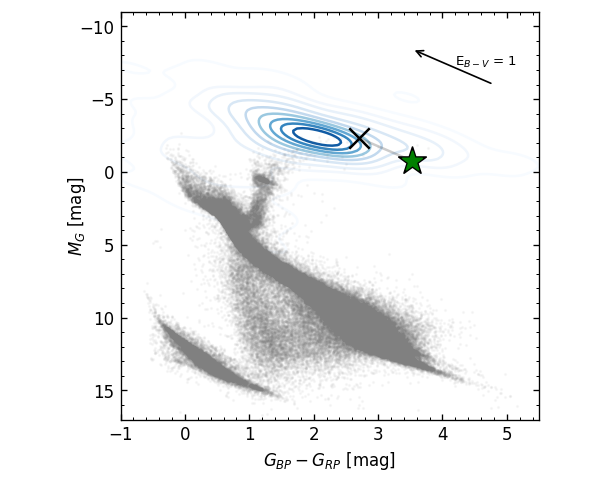}}
\caption{Position of V390 Sco in the \textit{Gaia} HR diagram. The green star symbol corresponds to the observed data, while the black cross symbol represents the magnitudes corrected for the reddening E$_{\rm (B-V)}=0.64$ mag, given by the map of \citet{2011ApJ...737..103S}. The absolute $G$ magnitude was calculated using the photogeometric distance from the catalog of \citet{2021AJ....161..147B}. The blue density map displays the position of known symbiotic variables from the New Online Database of Symbiotic Variables \citep{2019RNAAS...3...28M,2019AN....340..598M}. In gray, a sample of stars within 200 pc with reliable astrometry is shown for comparison \citep{2018A&A...616A..10G}.\label{fig:gaia_hr}}
\end{figure}

The light curves of V390 Sco exhibited variability inconsistent with that of a single Mira. The parallax measurement from \textit{Gaia}, despite its relative error of approximately 40\%, effectively rules out the possibility that V390 Sco is a main sequence star, considering the observed brightness of the star. The position of the target in the \textit{Gaia} color-magnitude diagram, adopting the distance of 5.3$\pm$0.8 kpc from the catalog of \citet{2021AJ....161..147B}, is characteristic of symbiotic stars (see Fig. \ref{fig:gaia_hr}). This evidence, coupled with the unexpected outburst-like behavior, prompted our spectroscopic follow-up to investigate its symbiotic nature.

We acquired 11 low-resolution spectra ($R \sim 500 - 3\,500$) of V390 Sco between September 28 and October 10, 2023 (JD\,2\,460\,215 - 2\,460\,227), after the maximum of the ongoing brightening when the star faded by about 0.5 and 0.4 mag in ATLAS $c$ and $o$ filters, respectively, in comparison with the outburst maximum. We utilized two remotely-controlled setups: a 30-cm Newton telescope equipped with Alpy600 spectrograph located in Chile and a 32-cm Planewave CDK telescope with UVEX spectrograph in Australia. In all cases, several (3-7) individual 20-minute exposures were stacked to obtain sufficient signal to noise. The log of observations is presented in Tab. \ref{tab:log}.

\begin{center}
\begin{table}[t]%
\centering
\caption{Log of spectroscopic observations.\label{tab:log}}%
\tabcolsep=0pt%
\begin{tabular*}{20pc}{@{\extracolsep\fill}cccc@{\extracolsep\fill}}
\toprule
\textbf{JD 2\,460\,..} & \textbf{Res.}  & \textbf{$\lambda_{\rm min}$-$\lambda_{\rm max}$}  & \textbf{Inst.}  \\
\midrule
215.57 & 554 & 3602-7808 & Newt. + Alpy600 \\
215.90 & 1475 & 4788-6633 & CDK + UVEX \\
215.99 & 1053 & 3749-5491 & CDK + UVEX \\
216.90 & 1143 & 5563-7431 & CDK + UVEX \\
216.99 & 1810 & 6881-8751 & CDK + UVEX \\
217.98 & 1042 & 7604-9459 & CDK + UVEX \\
219.91 & 1804 & 6881-8751 & CDK + UVEX \\
223.00 & 1506 & 4786-6633 & CDK + UVEX \\
224.96 & 3464 & 5978-6900 & CDK + UVEX \\
225.95 & 1950 & 4325-5238 & CDK + UVEX \\
227.95 & 1991 & 4325-5238 & CDK + UVEX \\
\bottomrule
\end{tabular*}
\end{table}
\end{center}

\begin{figure}[t]
\centerline{\includegraphics[width=\columnwidth]{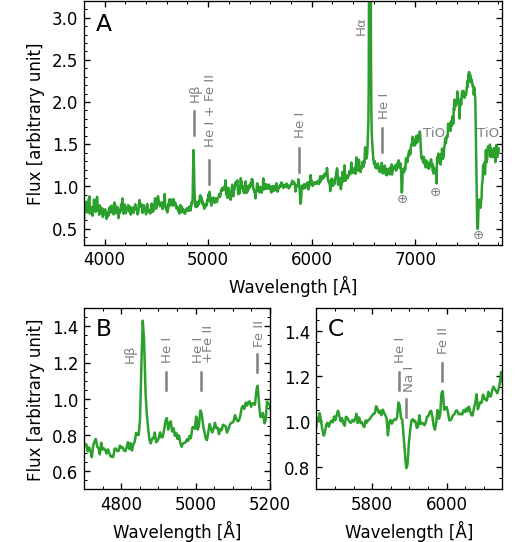}}
\caption{Optical low-resolution spectrum of V390 Sco acquired on September 28, 2023 (JD\,2\,460\,215). In panel \textbf{A}, the y-axis range is adjusted for the fainter lines, resulting in the H$\alpha$ line being partially cut off. Gray lines highlight the primary emission features, with much of the unlabeled emission corresponding to Fe II. Panels \textbf{B} and \textbf{C} provide close-ups of two specific regions of interest. \label{fig:spectra}}
\end{figure}

An example of the spectrum obtained on September 28, 2023 (JD\,2\,460\,215), is shown in Fig. \ref{fig:spectra}. No discernible changes are seen when compared with our other spectra obtained within the following two weeks. The observations revealed an M-type star continuum with strong emission lines of hydrogen and noticeable emission lines of He I and Fe II. Notably, no emission lines of higher ionization (e.g., [O III], He II, or [Fe VI]), usually seen in the quiescent spectra of symbiotic stars, were detected. The obtained data resemble a typical spectrum of a symbiotic star in an outburst when the expansion of the pseudo-atmosphere of the white dwarf leads to a decrease in the temperature of the ionizing source, resulting in lower ionization in the spectra \citep[e.g.,][and references therein]{2019arXiv190901389M}. Subsequently, during the transition from the outburst back to quiescence, the appearance of highly ionized emission lines is expected. In summary, considering both the spectroscopic appearance and photometric behavior of V390 Sco, we classify it as a symbiotic system.

The fit to the derredened spectrum \citep[E$_{\rm (B-V)}=0.64$ mag\footnote{A similar value of E$_{\rm (B-V)}=0.70$ was obtained using the \textit{mwdust} code \citep{2016ApJ...818..130B} employing the combined data from maps of \citet{2003A&A...409..205D}, \citet{2006A&A...453..635M}, and \citet{2019ApJ...887...93G}. OGLE-III data suggests E$_{\rm (B-V)}$ $\cong$ 0.625\,E$_{\rm (V-I)}$ = 0.54 in
the closeby line of sight \citep{2013ApJ...769...88N}.};][]{2011ApJ...737..103S} from Fig. \ref{fig:spectra} suggests the presence of an M6 giant and an additional source of radiation, which we approximated using a black-body spectrum with a temperature of 13\,000 K. The giant spectra were adopted from the MILES empirical library \citep[][]{2011A&A...532A..95F}. However, the spectral type of the red giant is not tightly constrained in this analysis. While a reasonably good fit was obtained, it depended on the parameters of the companion and the ratio between the sources, allowing for close spectral types to also provide acceptable fits. Nonetheless, the results are broadly consistent with those presented in Sec. \ref{sec:giant}.

The temperature of the hot source in the fit is also broadly consistent with the presence or absence of emission lines in the optical spectra. The lower limit can be estimated from the maximum ionisation potential $IP$ \citep[$T\rm_{h}$ $\sim$ $IP_{\text{max}}\times10^3$\,K;][]{1994A&A...282..586M}. Given the presence of Fe II lines but the absence of [O III] or He II, the temperature of the ionising source should be below a few $\times 10^4$\,K.

\section{Nature of the donor}\label{sec:giant}
As outlined in the Introduction, V390 Sco was previously classified as a Mira variable in the literature. The observational data collected and analyzed in this work strongly indicate that the target is a symbiotic system. Therefore, we further investigate in this section whether the donor in V390 Sco is a Mira-type star. Although most symbiotic binaries contain less evolved red giants, there exists a subgroup with Mira donors. Hence, the possibility of a Mira-type donor is not immediately excluded by confirming the symbiotic nature of the star.

Typically, symbiotic systems hosting Mira-type donors are categorized as D-type ("dusty") symbiotic stars, contrasting with S-type symbiotics ("stellar") that feature semiregularly pulsating giants. In D-type symbiotic stars, the highly evolved donor produces substantial amounts of dust, obscuring the signatures of the stellar photosphere in the optical spectrum \citep[e.g.,][]{2003ASPC..303...41W, 2012BaltA..21....5M}. This spectrum often resembles that of planetary nebulae, characterized by strong emission lines without a discernible continuum. Conversely, in S-type symbiotic stars, the presence of the red giant is easily distinguishable in the optical spectrum thanks to prominent absorption bands. 

\begin{figure}[t]
\centerline{\includegraphics[width=0.82\columnwidth]{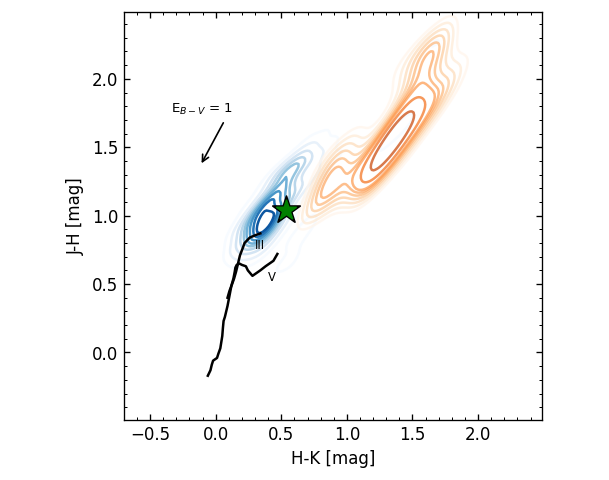}}
\caption{Position of V390 Sco in the 2MASS color-color diagram (depicted by the green star symbol). The blue and orange density map illustrates the locations of known S- and D-type symbiotic variables from the New Online Database of Symbiotic Variables \citep{2019RNAAS...3...28M,2019AN....340..598M}. Black solid lines represent sequences for main-sequence and giant stars \citep{2009BaltA..18...19S}.\label{fig:ir}}
\end{figure}

The optical spectra obtained for V390 Sco indicate an S-type classification, a conclusion supported by the position of the star in the infrared color-color diagram (Fig. \ref{fig:ir}). In this diagram, V390 Sco occupies the region associated with S-type symbiotic stars rather than systems containing Mira-type donors. Additionally, the IR criteria established by \citet{2019MNRAS.483.5077A} do not categorize the star as a D-type symbiotic star either. 

Distinguishing between these types based on IR data is possible due to the presence of hot dust \citep[with temperatures around $\sim$700 to 1\,000 K; e.g.,][]{1982ASSL...95...27A,2010MNRAS.402.2075A} in symbiotic stars with Mira donors, leading to a distinct infrared excess. In these systems, the spectral energy distribution (SED) exhibits a peak at longer wavelengths, approximately 2 to 4 µm, in contrast to S-type symbiotic stars where the peak occurs at 0.8 to 1.7 µm \citep{1995MNRAS.273..517I, 2019ApJS..240...21A}. The SED of V390 Sco aligns with the characteristics of an S-type symbiotic system. The slight overflux is evident only in WISE data. However, the system was in an outburst during the observations cataloged in the AllWISE database (Sec. \ref{sec:photometry}), potentially accounting for the marginal excess observed in the SED. It is important to note that flags in the database indicate potential contamination of (at least) the $W1$ and $W2$ observations by nearby bright sources, raising concerns about the reliability of the photometry. Consequently, we refrain from drawing further conclusions from the WISE photometry in this context. In any case, the excess is not as prominent as usually observed in D-type symbiotic stars.

\begin{figure}[t]
\centerline{\includegraphics[width=\columnwidth]{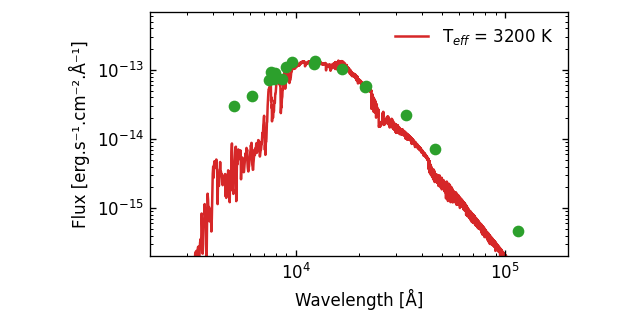}}
\caption{SED of V390 Sco (depicted by green points), compiled from archival observations from \textit{Gaia} DR3, 2MASS, WISE, and DENIS surveys \citep{1997Msngr..87...27E}. The VOSA service was utilized for data retrieval, and corrections for reddening of E$_{\rm (B-V)}=0.64$ mag were applied \citep{2011ApJ...737..103S}. A BT-Settl theoretical spectrum of a red giant with an effective temperature of 3\,200 K is presented in red for comparison \citep{2014IAUS..299..271A}. The blue excess is attributed to the additional radiation components (hot component, symbiotic nebula) and the fact that \textit{Gaia} data were acquired mostly during the outburst. The infrared excess is discussed in detail in the main text. \label{fig:sed}}
\end{figure}

The photometric observations used to construct the SED in Fig. \ref{fig:sed} were not simultaneous, making it challenging to accurately estimate the parameters of the components of the system. Additionally, the measured fluxes represent not only the radiation from the donor but also contributions from the hot component and symbiotic nebula, with these contributions varying during outbursts and in quiescence. Despite these complexities, attempting to fit the SED with a giant template spectrum having a reasonable effective temperature (approximately $2800 - 4000$ K, typical for symbiotic giants), assuming a distance of 5.3 kpc \citep{2021AJ....161..147B}, and allowing for UV and IR excess consistently yielded a giant luminosity of only a few $\times 10^1$\,L$\odot$. This luminosity is significantly lower than expected for an evolved asymptotic giant branch star pulsating as a Mira variable. The analysis was conducted using the VO SED analyzer (VOSA) on the Spanish Virtual Observatory theoretical services website \citep{2008A&A...492..277B}\footnote{http://svo2.cab.inta-csic.es/theory/vosa/}.

Finally, we revisit the photometric time series observations. As mentioned in Sec. \ref{sec:photometry}, we identified periodic variability in the ATLAS, OGLE, and \textit{Gaia} light curves of V390 Sco. However, at first glance, this data does not resemble the pulsations of a Mira variable, which are typically observed with much higher amplitude. We note here that while high-amplitude variability in single Miras is usually detectable across all wavelengths, in symbiotic systems, the contribution of other sources at optical wavelengths (hot component and/or nebula) or the dust obscuration may complicate the detection of Mira pulsations. As an illustrative example, we refer the reader to the case of V1016 Cyg, a D-type symbiotic star in which Mira pulsations are not visible in $U$, $B$, $V$, and $R_{\rm c}$ filters, while they are very prominent in the $I_{\rm c}$ filter \citep[see, e.g., fig. 13 in][]{2019CoSka..49...19S}. Other similar examples include RX Pup and V366 Car \citep{2009AcA....59..169G}, as well as [JD2002] 11 \citep{2015AcA....65..139H}.

To support the conclusion that the photometric data does not suggest the presence of a Mira pulsator, we compared the variability observed in V390 Sco in \textit{Gaia} and ATLAS with the variability observed in other symbiotic stars using the same datasets. \textit{Gaia} data are particularly useful in this context, considering the wavelength-dependent nature of pulsation detection mentioned above, given the very broadband \textit{Gaia} filters. The detected amplitude of variability in V390 Sco is approximately 0.2 mag and 0.4 mag in \textit{Gaia} $G$ and ATLAS $o$, respectively. This is significantly lower compared to other symbiotic stars with Mira donors from the New Online Database of Symbiotic Variables. The average amplitude in Gaia $G$ for a sample of 20 D-type symbiotic stars is approximately 1 mag, about five times higher than the case of V390 Sco. On the other hand, S-type symbiotic stars hosting semiregularly pulsating giants exhibit variability at the level of about 0.1 - 0.2 mag in \textit{Gaia} $G$. In ATLAS, some D-type symbiotic stars, such as EF Aql or LL Cas, show variability of up to 3.5 mag in $o$. 

In the case of S-type symbiotic stars, variability is not necessarily solely attributed to pulsations; in some cases, the periodicity arises from the orbital motion of the binary. However, the 190-day period observed in V390 Sco appears to be on the lower end of the range of orbital periods in symbiotic stars. Most S-type symbiotic binaries exhibit periods in the range of 300 to 1\,100 days, with a peak between 500 and 600 days \citep[e.g.,][]{1999A&AS..137..473M,2012BaltA..21....5M}. Conversely, the pulsation periods of S-type symbiotic stars typically cluster around 50 to 200 days \citep[e.g.,][]{2013AcA....63..405G}. 

In conclusion, the photometric variability, optical spectroscopic data, infrared observations, and the SED all support the S-type classification of V390 Sco and do not indicate the presence of a Mira variable in the system. Additional insights into the nature of the observed period in photometry, whether linked to orbital motion or the pulsation of the donor, may be gained through multicolor photometric observations or radial velocity studies.

\section{Conclusions}\label{sec:conclusions}
The variable star V390 Sco, previously classified as a Mira pulsator, exhibited a brightening event in September 2023 detected by $Gaia$ satellite. This behavior, reported through the GSA, unusual for a single Mira star, prompted our spectroscopic follow-up. In this study, we comprehensively analyzed archival multifrequency photometric data, including \textit{Gaia}, ASAS, ASAS-SN, OGLE, and ATLAS observations, and presented our spectroscopic results.

Our analysis supports the classification of V390 Sco as a symbiotic star. The photometric data revealed not only the recent brightening but also three additional outbursts (in 2004, 2010, and 2017), consistent with the characteristic behavior of classical symbiotic stars. The outbursts recur approximately every 2330 -- 2400 days, and we hypothesize the periastron passage in an eccentric orbit may trigger them, similarly to the case of BX Mon, DD Mic, or MWC~560. We also investigated whether V390 Sco hosts a Mira pulsator. Our findings suggest that it is more likely an S-type symbiotic system containing a red giant that is less evolved than a typical~Mira.

V390 Sco (Gaia23ckh) adds to the growing list of symbiotic stars discovered through the GSA system. Notably, \citet{2020A&A...644A..49M} previously identified Gaia18aen as a new symbiotic system, marking the first discovery of a binary star of this class by \textit{Gaia}. Furthermore, \citet{2022PhDT........34M} classified Gaia17bjg as a classical symbiotic star in M31, signifying the first confirmation of an extragalactic symbiotic binary by \textit{Gaia}. 

The observation of \textit{Gaia} also led to the detection of the first recorded outburst of the symbiotic star WRAY 15-1167 \citep[Gaia22bou; ][]{2022ATel15340....1M}, and the detection of the fading of V2756 Sgr by \textit{Gaia} (Gaia22eor) led to the revelation of its eclipsing nature \citep[][]{2022RNAAS...6..253M}. While the current work does not categorize V390 Sco as the first D-type symbiotic star discovered by \textit{Gaia}, it emphasizes the ongoing importance of \textit{Gaia} in advancing our understanding of the dynamic behavior exhibited by symbiotic stars.


\section*{Acknowledgments}
We are thankful to an anonymous referee for the comments and suggestions improving the manuscript. The research of JM was supported by project \fundingNumber{COOPERATIO - PHYSICS} of \fundingAgency{Charles University in Prague}.

This work has made use of data from the Asteroid Terrestrial-impact Last Alert System (ATLAS) project. The Asteroid Terrestrial-impact Last Alert System (ATLAS) project is primarily funded to search for near earth asteroids through NASA grants NN12AR55G, 80NSSC18K0284, and 80NSSC18K1575; byproducts of the NEO search include images and catalogs from the survey area. This work was partially funded by Kepler/K2 grant J1944/80NSSC19K0112 and HST GO-15889, and STFC grants ST/T000198/1 and ST/S006109/1. The ATLAS science products have been made possible through the contributions of the University of Hawaii Institute for Astronomy, the Queen’s University Belfast, the Space Telescope Science Institute, the South African Astronomical Observatory, and The Millennium Institute of Astrophysics (MAS), Chile.





\section*{Conflict of interest statement}

The authors declare no potential conflict of interests.

\bibliography{Wiley-ASNA}%

\begin{thebibliography}{}

\bibitem [\protect \citeauthoryear {%
{Akras}%
, {Guzman-Ramirez}%
, {Leal-Ferreira}%
\BCBL {}\ \BBA {} {Ramos-Larios}%
}{%
{Akras}%
, {Guzman-Ramirez}%
\BCBL {}\ \protect \BOthers {.}}{%
{\protect \APACyear {2019}}%
}]{%
2019ApJS..240...21A}
\APACinsertmetastar {%
2019ApJS..240...21A}%
\begin{APACrefauthors}%
{Akras}, S.%
, {Guzman-Ramirez}, L.%
, {Leal-Ferreira}, M\BPBI L.%
\BCBL {}\ \BBA {} {Ramos-Larios}, G.%
\end{APACrefauthors}%
\unskip\
\newblock
\APACrefYearMonthDay{2019}{{\APACmonth{02}}}{},
\newblock
\unskip
\newblock
\APACjournalVolNumPages{\apjs}{240}{2}{21}.
\newblock
\begin{APACrefDOI} \doi{10.3847/1538-4365/aaf88c} \end{APACrefDOI}
\PrintBackRefs{\CurrentBib}

\bibitem [\protect \citeauthoryear {%
{Akras}%
, {Leal-Ferreira}%
, {Guzman-Ramirez}%
\BCBL {}\ \BBA {} {Ramos-Larios}%
}{%
{Akras}%
, {Leal-Ferreira}%
\BCBL {}\ \protect \BOthers {.}}{%
{\protect \APACyear {2019}}%
}]{%
2019MNRAS.483.5077A}
\APACinsertmetastar {%
2019MNRAS.483.5077A}%
\begin{APACrefauthors}%
{Akras}, S.%
, {Leal-Ferreira}, M\BPBI L.%
, {Guzman-Ramirez}, L.%
\BCBL {}\ \BBA {} {Ramos-Larios}, G.%
\end{APACrefauthors}%
\unskip\
\newblock
\APACrefYearMonthDay{2019}{{\APACmonth{03}}}{},
\newblock
\unskip
\newblock
\APACjournalVolNumPages{\mnras}{483}{4}{5077-5104}.
\newblock
\begin{APACrefDOI} \doi{10.1093/mnras/sty3359} \end{APACrefDOI}
\PrintBackRefs{\CurrentBib}

\bibitem [\protect \citeauthoryear {%
{Allard}%
}{%
{Allard}%
}{%
{\protect \APACyear {2014}}%
}]{%
2014IAUS..299..271A}
\APACinsertmetastar {%
2014IAUS..299..271A}%
\begin{APACrefauthors}%
{Allard}, F.%
\end{APACrefauthors}%
\unskip\
\newblock
\APACrefYearMonthDay{2014}{{\APACmonth{01}}}{},
\newblock
{\BBOQ}\APACrefatitle {{The BT-Settl Model Atmospheres for Stars, Brown Dwarfs and Planets}} {{The BT-Settl Model Atmospheres for Stars, Brown Dwarfs and Planets}}.{\BBCQ}
\newblock
\BIn{} M.~{Booth}, B\BPBI C.~{Matthews}\BCBL {}\ \BBA {} J\BPBI R.~{Graham}\ (\BEDS), \APACrefbtitle {Exploring the Formation and Evolution of Planetary Systems} {Exploring the Formation and Evolution of Planetary Systems}\ \BVOL~299, \BPG~271-272.
\newblock
\begin{APACrefDOI} \doi{10.1017/S1743921313008545} \end{APACrefDOI}
\PrintBackRefs{\CurrentBib}

\bibitem [\protect \citeauthoryear {%
{Allen}%
}{%
{Allen}%
}{%
{\protect \APACyear {1982}}%
}]{%
1982ASSL...95...27A}
\APACinsertmetastar {%
1982ASSL...95...27A}%
\begin{APACrefauthors}%
{Allen}, D\BPBI A.%
\end{APACrefauthors}%
\unskip\
\newblock
\APACrefYearMonthDay{1982}{{\APACmonth{09}}}{},
\newblock
{\BBOQ}\APACrefatitle {{Infrared studies of symbiotic stars.}} {{Infrared studies of symbiotic stars.}}{\BBCQ}
\newblock
\BIn{} M.~{Friedjung}\ \BBA {} R.~{Viotti}\ (\BEDS), \APACrefbtitle {IAU Colloq. 70: The Nature of Symbiotic Stars} {IAU Colloq. 70: The Nature of Symbiotic Stars}\ \BVOL~95, \BPG~27-42.
\newblock
\begin{APACrefDOI} \doi{10.1007/978-94-009-7834-8_4} \end{APACrefDOI}
\PrintBackRefs{\CurrentBib}

\bibitem [\protect \citeauthoryear {%
{Angeloni}%
, {Contini}%
, {Ciroi}%
\BCBL {}\ \BBA {} {Rafanelli}%
}{%
{Angeloni}%
\ \protect \BOthers {.}}{%
{\protect \APACyear {2010}}%
}]{%
2010MNRAS.402.2075A}
\APACinsertmetastar {%
2010MNRAS.402.2075A}%
\begin{APACrefauthors}%
{Angeloni}, R.%
, {Contini}, M.%
, {Ciroi}, S.%
\BCBL {}\ \BBA {} {Rafanelli}, P.%
\end{APACrefauthors}%
\unskip\
\newblock
\APACrefYearMonthDay{2010}{{\APACmonth{03}}}{},
\newblock
\unskip
\newblock
\APACjournalVolNumPages{\mnras}{402}{3}{2075-2086}.
\newblock
\begin{APACrefDOI} \doi{10.1111/j.1365-2966.2009.16067.x} \end{APACrefDOI}
\PrintBackRefs{\CurrentBib}

\bibitem [\protect \citeauthoryear {%
{Bailer-Jones}%
, {Rybizki}%
, {Fouesneau}%
, {Demleitner}%
\BCBL {}\ \BBA {} {Andrae}%
}{%
{Bailer-Jones}%
\ \protect \BOthers {.}}{%
{\protect \APACyear {2021}}%
}]{%
2021AJ....161..147B}
\APACinsertmetastar {%
2021AJ....161..147B}%
\begin{APACrefauthors}%
{Bailer-Jones}, C\BPBI A\BPBI L.%
, {Rybizki}, J.%
, {Fouesneau}, M.%
, {Demleitner}, M.%
\BCBL {}\ \BBA {} {Andrae}, R.%
\end{APACrefauthors}%
\unskip\
\newblock
\APACrefYearMonthDay{2021}{{\APACmonth{03}}}{},
\newblock
\unskip
\newblock
\APACjournalVolNumPages{\aj}{161}{3}{147}.
\newblock
\begin{APACrefDOI} \doi{10.3847/1538-3881/abd806} \end{APACrefDOI}
\PrintBackRefs{\CurrentBib}

\bibitem [\protect \citeauthoryear {%
{Bayo}%
\ \protect \BOthers {.}}{%
{Bayo}%
\ \protect \BOthers {.}}{%
{\protect \APACyear {2008}}%
}]{%
2008A&A...492..277B}
\APACinsertmetastar {%
2008A&A...492..277B}%
\begin{APACrefauthors}%
{Bayo}, A.%
, {Rodrigo}, C.%
, {Barrado Y Navascu{\'e}s}, D.%
, {Solano}, E.%
, {Guti{\'e}rrez}, R.%
, {Morales-Calder{\'o}n}, M.%
\BCBL {}\ \BBA {} {Allard}, F.%
\end{APACrefauthors}%
\unskip\
\newblock
\APACrefYearMonthDay{2008}{{\APACmonth{12}}}{},
\newblock
\unskip
\newblock
\APACjournalVolNumPages{\aap}{492}{1}{277-287}.
\newblock
\begin{APACrefDOI} \doi{10.1051/0004-6361:200810395} \end{APACrefDOI}
\PrintBackRefs{\CurrentBib}

\bibitem [\protect \citeauthoryear {%
{Bovy}%
, {Rix}%
, {Green}%
, {Schlafly}%
\BCBL {}\ \BBA {} {Finkbeiner}%
}{%
{Bovy}%
\ \protect \BOthers {.}}{%
{\protect \APACyear {2016}}%
}]{%
2016ApJ...818..130B}
\APACinsertmetastar {%
2016ApJ...818..130B}%
\begin{APACrefauthors}%
{Bovy}, J.%
, {Rix}, H\BHBI W.%
, {Green}, G\BPBI M.%
, {Schlafly}, E\BPBI F.%
\BCBL {}\ \BBA {} {Finkbeiner}, D\BPBI P.%
\end{APACrefauthors}%
\unskip\
\newblock
\APACrefYearMonthDay{2016}{{\APACmonth{02}}}{},
\newblock
\unskip
\newblock
\APACjournalVolNumPages{\apj}{818}{2}{130}.
\newblock
\begin{APACrefDOI} \doi{10.3847/0004-637X/818/2/130} \end{APACrefDOI}
\PrintBackRefs{\CurrentBib}

\bibitem [\protect \citeauthoryear {%
{Drimmel}%
, {Cabrera-Lavers}%
\BCBL {}\ \BBA {} {L{\'o}pez-Corredoira}%
}{%
{Drimmel}%
\ \protect \BOthers {.}}{%
{\protect \APACyear {2003}}%
}]{%
2003A&A...409..205D}
\APACinsertmetastar {%
2003A&A...409..205D}%
\begin{APACrefauthors}%
{Drimmel}, R.%
, {Cabrera-Lavers}, A.%
\BCBL {}\ \BBA {} {L{\'o}pez-Corredoira}, M.%
\end{APACrefauthors}%
\unskip\
\newblock
\APACrefYearMonthDay{2003}{{\APACmonth{10}}}{},
\newblock
\unskip
\newblock
\APACjournalVolNumPages{\aap}{409}{}{205-215}.
\newblock
\begin{APACrefDOI} \doi{10.1051/0004-6361:20031070} \end{APACrefDOI}
\PrintBackRefs{\CurrentBib}

\bibitem [\protect \citeauthoryear {%
{Epchtein}%
\ \protect \BOthers {.}}{%
{Epchtein}%
\ \protect \BOthers {.}}{%
{\protect \APACyear {1997}}%
}]{%
1997Msngr..87...27E}
\APACinsertmetastar {%
1997Msngr..87...27E}%
\begin{APACrefauthors}%
{Epchtein}, N.%
, {de Batz}, B.%
, {Capoani}, L.%
\ et al.\end{APACrefauthors}%
\unskip\
\newblock
\APACrefYearMonthDay{1997}{{\APACmonth{03}}}{},
\newblock
\unskip
\newblock
\APACjournalVolNumPages{The Messenger}{87}{}{27-34}.
\PrintBackRefs{\CurrentBib}

\bibitem [\protect \citeauthoryear {%
{Falc{\'o}n-Barroso}%
\ \protect \BOthers {.}}{%
{Falc{\'o}n-Barroso}%
\ \protect \BOthers {.}}{%
{\protect \APACyear {2011}}%
}]{%
2011A&A...532A..95F}
\APACinsertmetastar {%
2011A&A...532A..95F}%
\begin{APACrefauthors}%
{Falc{\'o}n-Barroso}, J.%
, {S{\'a}nchez-Bl{\'a}zquez}, P.%
, {Vazdekis}, A.%
\ et al.\end{APACrefauthors}%
\unskip\
\newblock
\APACrefYearMonthDay{2011}{{\APACmonth{08}}}{},
\newblock
\unskip
\newblock
\APACjournalVolNumPages{\aap}{532}{}{A95}.
\newblock
\begin{APACrefDOI} \doi{10.1051/0004-6361/201116842} \end{APACrefDOI}
\PrintBackRefs{\CurrentBib}

\bibitem [\protect \citeauthoryear {%
{Gaia Collaboration}%
\ \protect \BOthers {.}}{%
{Gaia Collaboration}%
\ \protect \BOthers {.}}{%
{\protect \APACyear {2018}}%
}]{%
2018A&A...616A..10G}
\APACinsertmetastar {%
2018A&A...616A..10G}%
\begin{APACrefauthors}%
{Gaia Collaboration}%
, {Babusiaux}, C.%
, {van Leeuwen}, F.%
\ et al.\end{APACrefauthors}%
\unskip\
\newblock
\APACrefYearMonthDay{2018}{{\APACmonth{08}}}{},
\newblock
\unskip
\newblock
\APACjournalVolNumPages{\aap}{616}{}{A10}.
\newblock
\begin{APACrefDOI} \doi{10.1051/0004-6361/201832843} \end{APACrefDOI}
\PrintBackRefs{\CurrentBib}

\bibitem [\protect \citeauthoryear {%
{Gaia Collaboration}%
\ \protect \BOthers {.}}{%
{Gaia Collaboration}%
\ \protect \BOthers {.}}{%
{\protect \APACyear {2016}}%
}]{%
2016A&A...595A...1G}
\APACinsertmetastar {%
2016A&A...595A...1G}%
\begin{APACrefauthors}%
{Gaia Collaboration}%
, {Prusti}, T.%
, {de Bruijne}, J\BPBI H\BPBI J.%
\ et al.\end{APACrefauthors}%
\unskip\
\newblock
\APACrefYearMonthDay{2016}{{\APACmonth{11}}}{},
\newblock
\unskip
\newblock
\APACjournalVolNumPages{\aap}{595}{}{A1}.
\newblock
\begin{APACrefDOI} \doi{10.1051/0004-6361/201629272} \end{APACrefDOI}
\PrintBackRefs{\CurrentBib}

\bibitem [\protect \citeauthoryear {%
{Gaia Collaboration}%
\ \protect \BOthers {.}}{%
{Gaia Collaboration}%
\ \protect \BOthers {.}}{%
{\protect \APACyear {2023}}%
}]{%
2023A&A...674A...1G}
\APACinsertmetastar {%
2023A&A...674A...1G}%
\begin{APACrefauthors}%
{Gaia Collaboration}%
, {Vallenari}, A.%
, {Brown}, A\BPBI G\BPBI A.%
\ et al.\end{APACrefauthors}%
\unskip\
\newblock
\APACrefYearMonthDay{2023}{{\APACmonth{06}}}{},
\newblock
\unskip
\newblock
\APACjournalVolNumPages{\aap}{674}{}{A1}.
\newblock
\begin{APACrefDOI} \doi{10.1051/0004-6361/202243940} \end{APACrefDOI}
\PrintBackRefs{\CurrentBib}

\bibitem [\protect \citeauthoryear {%
{Green}%
, {Schlafly}%
, {Zucker}%
, {Speagle}%
\BCBL {}\ \BBA {} {Finkbeiner}%
}{%
{Green}%
\ \protect \BOthers {.}}{%
{\protect \APACyear {2019}}%
}]{%
2019ApJ...887...93G}
\APACinsertmetastar {%
2019ApJ...887...93G}%
\begin{APACrefauthors}%
{Green}, G\BPBI M.%
, {Schlafly}, E.%
, {Zucker}, C.%
, {Speagle}, J\BPBI S.%
\BCBL {}\ \BBA {} {Finkbeiner}, D.%
\end{APACrefauthors}%
\unskip\
\newblock
\APACrefYearMonthDay{2019}{{\APACmonth{12}}}{},
\newblock
\unskip
\newblock
\APACjournalVolNumPages{\apj}{887}{1}{93}.
\newblock
\begin{APACrefDOI} \doi{10.3847/1538-4357/ab5362} \end{APACrefDOI}
\PrintBackRefs{\CurrentBib}

\bibitem [\protect \citeauthoryear {%
{Gromadzki}%
, {Miko{\l}ajewska}%
\BCBL {}\ \BBA {} {Soszy{\'n}ski}%
}{%
{Gromadzki}%
\ \protect \BOthers {.}}{%
{\protect \APACyear {2013}}%
}]{%
2013AcA....63..405G}
\APACinsertmetastar {%
2013AcA....63..405G}%
\begin{APACrefauthors}%
{Gromadzki}, M.%
, {Miko{\l}ajewska}, J.%
\BCBL {}\ \BBA {} {Soszy{\'n}ski}, I.%
\end{APACrefauthors}%
\unskip\
\newblock
\APACrefYearMonthDay{2013}{{\APACmonth{12}}}{},
\newblock
\unskip
\newblock
\APACjournalVolNumPages{\actaa}{63}{4}{405-428}.
\newblock
\begin{APACrefDOI} \doi{10.48550/arXiv.1312.6063} \end{APACrefDOI}
\PrintBackRefs{\CurrentBib}

\bibitem [\protect \citeauthoryear {%
{Gromadzki}%
, {Miko{\l}ajewska}%
, {Whitelock}%
\BCBL {}\ \BBA {} {Marang}%
}{%
{Gromadzki}%
\ \protect \BOthers {.}}{%
{\protect \APACyear {2009}}%
}]{%
2009AcA....59..169G}
\APACinsertmetastar {%
2009AcA....59..169G}%
\begin{APACrefauthors}%
{Gromadzki}, M.%
, {Miko{\l}ajewska}, J.%
, {Whitelock}, P.%
\BCBL {}\ \BBA {} {Marang}, F.%
\end{APACrefauthors}%
\unskip\
\newblock
\APACrefYearMonthDay{2009}{{\APACmonth{06}}}{},
\newblock
\unskip
\newblock
\APACjournalVolNumPages{\actaa}{59}{2}{169-191}.
\newblock
\begin{APACrefDOI} \doi{10.48550/arXiv.0906.4136} \end{APACrefDOI}
\PrintBackRefs{\CurrentBib}

\bibitem [\protect \citeauthoryear {%
{Gromadzki}%
, {Miko{\l}ajewska}%
, {Whitelock}%
\BCBL {}\ \BBA {} {Marang}%
}{%
{Gromadzki}%
\ \protect \BOthers {.}}{%
{\protect \APACyear {2007}}%
}]{%
2007A&A...463..703G}
\APACinsertmetastar {%
2007A&A...463..703G}%
\begin{APACrefauthors}%
{Gromadzki}, M.%
, {Miko{\l}ajewska}, J.%
, {Whitelock}, P\BPBI A.%
\BCBL {}\ \BBA {} {Marang}, F.%
\end{APACrefauthors}%
\unskip\
\newblock
\APACrefYearMonthDay{2007}{{\APACmonth{02}}}{},
\newblock
\unskip
\newblock
\APACjournalVolNumPages{\aap}{463}{2}{703-706}.
\newblock
\begin{APACrefDOI} \doi{10.1051/0004-6361:20066538} \end{APACrefDOI}
\PrintBackRefs{\CurrentBib}

\bibitem [\protect \citeauthoryear {%
{Hajduk}%
, {Gromadzki}%
, {Miko{\l}ajewska}%
, {Miszalski}%
\BCBL {}\ \BBA {} {Soszy{\'n}ski}%
}{%
{Hajduk}%
\ \protect \BOthers {.}}{%
{\protect \APACyear {2015}}%
}]{%
2015AcA....65..139H}
\APACinsertmetastar {%
2015AcA....65..139H}%
\begin{APACrefauthors}%
{Hajduk}, M.%
, {Gromadzki}, M.%
, {Miko{\l}ajewska}, J.%
, {Miszalski}, B.%
\BCBL {}\ \BBA {} {Soszy{\'n}ski}, I.%
\end{APACrefauthors}%
\unskip\
\newblock
\APACrefYearMonthDay{2015}{{\APACmonth{06}}}{},
\newblock
\unskip
\newblock
\APACjournalVolNumPages{\actaa}{65}{2}{139-149}.
\newblock
\begin{APACrefDOI} \doi{10.48550/arXiv.1508.00089} \end{APACrefDOI}
\PrintBackRefs{\CurrentBib}

\bibitem [\protect \citeauthoryear {%
{Hodgkin}%
\ \protect \BOthers {.}}{%
{Hodgkin}%
\ \protect \BOthers {.}}{%
{\protect \APACyear {2023}}%
}]{%
2023TNSTR2302....1H}
\APACinsertmetastar {%
2023TNSTR2302....1H}%
\begin{APACrefauthors}%
{Hodgkin}, S\BPBI T.%
, {Breedt}, E.%
, {Delgado}, A.%
\ et al.\end{APACrefauthors}%
\unskip\
\newblock
\APACrefYearMonthDay{2023}{{\APACmonth{09}}}{},
\newblock
\unskip
\newblock
\APACjournalVolNumPages{Transient Name Server Discovery Report}{2023-2302}{}{1}.
\PrintBackRefs{\CurrentBib}

\bibitem [\protect \citeauthoryear {%
{Hodgkin}%
\ \protect \BOthers {.}}{%
{Hodgkin}%
\ \protect \BOthers {.}}{%
{\protect \APACyear {2021}}%
}]{%
2021A&A...652A..76H}
\APACinsertmetastar {%
2021A&A...652A..76H}%
\begin{APACrefauthors}%
{Hodgkin}, S\BPBI T.%
, {Harrison}, D\BPBI L.%
, {Breedt}, E.%
\ et al.\end{APACrefauthors}%
\unskip\
\newblock
\APACrefYearMonthDay{2021}{{\APACmonth{08}}}{},
\newblock
\unskip
\newblock
\APACjournalVolNumPages{\aap}{652}{}{A76}.
\newblock
\begin{APACrefDOI} \doi{10.1051/0004-6361/202140735} \end{APACrefDOI}
\PrintBackRefs{\CurrentBib}

\bibitem [\protect \citeauthoryear {%
{Ivison}%
, {Seaquist}%
, {Schwarz}%
, {Hughes}%
\BCBL {}\ \BBA {} {Bode}%
}{%
{Ivison}%
\ \protect \BOthers {.}}{%
{\protect \APACyear {1995}}%
}]{%
1995MNRAS.273..517I}
\APACinsertmetastar {%
1995MNRAS.273..517I}%
\begin{APACrefauthors}%
{Ivison}, R\BPBI J.%
, {Seaquist}, E\BPBI R.%
, {Schwarz}, H\BPBI E.%
, {Hughes}, D\BPBI H.%
\BCBL {}\ \BBA {} {Bode}, M\BPBI F.%
\end{APACrefauthors}%
\unskip\
\newblock
\APACrefYearMonthDay{1995}{{\APACmonth{03}}}{},
\newblock
\unskip
\newblock
\APACjournalVolNumPages{\mnras}{273}{2}{517-527}.
\newblock
\begin{APACrefDOI} \doi{10.1093/mnras/273.2.517} \end{APACrefDOI}
\PrintBackRefs{\CurrentBib}

\bibitem [\protect \citeauthoryear {%
{Kochanek}%
\ \protect \BOthers {.}}{%
{Kochanek}%
\ \protect \BOthers {.}}{%
{\protect \APACyear {2017}}%
}]{%
2017PASP..129j4502K}
\APACinsertmetastar {%
2017PASP..129j4502K}%
\begin{APACrefauthors}%
{Kochanek}, C\BPBI S.%
, {Shappee}, B\BPBI J.%
, {Stanek}, K\BPBI Z.%
\ et al.\end{APACrefauthors}%
\unskip\
\newblock
\APACrefYearMonthDay{2017}{{\APACmonth{10}}}{},
\newblock
\unskip
\newblock
\APACjournalVolNumPages{\pasp}{129}{980}{104502}.
\newblock
\begin{APACrefDOI} \doi{10.1088/1538-3873/aa80d9} \end{APACrefDOI}
\PrintBackRefs{\CurrentBib}

\bibitem [\protect \citeauthoryear {%
{Lebzelter}%
\ \protect \BOthers {.}}{%
{Lebzelter}%
\ \protect \BOthers {.}}{%
{\protect \APACyear {2023}}%
}]{%
2023A&A...674A..15L}
\APACinsertmetastar {%
2023A&A...674A..15L}%
\begin{APACrefauthors}%
{Lebzelter}, T.%
, {Mowlavi}, N.%
, {Lecoeur-Taibi}, I.%
\ et al.\end{APACrefauthors}%
\unskip\
\newblock
\APACrefYearMonthDay{2023}{{\APACmonth{06}}}{},
\newblock
\unskip
\newblock
\APACjournalVolNumPages{\aap}{674}{}{A15}.
\newblock
\begin{APACrefDOI} \doi{10.1051/0004-6361/202244241} \end{APACrefDOI}
\PrintBackRefs{\CurrentBib}

\bibitem [\protect \citeauthoryear {%
{Lomb}%
}{%
{Lomb}%
}{%
{\protect \APACyear {1976}}%
}]{%
1976Ap&SS..39..447L}
\APACinsertmetastar {%
1976Ap&SS..39..447L}%
\begin{APACrefauthors}%
{Lomb}, N\BPBI R.%
\end{APACrefauthors}%
\unskip\
\newblock
\APACrefYearMonthDay{1976}{{\APACmonth{02}}}{},
\newblock
\unskip
\newblock
\APACjournalVolNumPages{\apss}{39}{2}{447-462}.
\newblock
\begin{APACrefDOI} \doi{10.1007/BF00648343} \end{APACrefDOI}
\PrintBackRefs{\CurrentBib}

\bibitem [\protect \citeauthoryear {%
{Magdolen}%
\ \protect \BOthers {.}}{%
{Magdolen}%
\ \protect \BOthers {.}}{%
{\protect \APACyear {2023}}%
}]{%
2023A&A...675A.140M}
\APACinsertmetastar {%
2023A&A...675A.140M}%
\begin{APACrefauthors}%
{Magdolen}, J.%
, {Dobrotka}, A.%
, {Orio}, M.%
\ et al.\end{APACrefauthors}%
\unskip\
\newblock
\APACrefYearMonthDay{2023}{{\APACmonth{07}}}{},
\newblock
\unskip
\newblock
\APACjournalVolNumPages{\aap}{675}{}{A140}.
\newblock
\begin{APACrefDOI} \doi{10.1051/0004-6361/202345935} \end{APACrefDOI}
\PrintBackRefs{\CurrentBib}

\bibitem [\protect \citeauthoryear {%
{Mainzer}%
\ \protect \BOthers {.}}{%
{Mainzer}%
\ \protect \BOthers {.}}{%
{\protect \APACyear {2014}}%
}]{%
2014ApJ...792...30M}
\APACinsertmetastar {%
2014ApJ...792...30M}%
\begin{APACrefauthors}%
{Mainzer}, A.%
, {Bauer}, J.%
, {Cutri}, R\BPBI M.%
\ et al.\end{APACrefauthors}%
\unskip\
\newblock
\APACrefYearMonthDay{2014}{{\APACmonth{09}}}{},
\newblock
\unskip
\newblock
\APACjournalVolNumPages{\apj}{792}{1}{30}.
\newblock
\begin{APACrefDOI} \doi{10.1088/0004-637X/792/1/30} \end{APACrefDOI}
\PrintBackRefs{\CurrentBib}

\bibitem [\protect \citeauthoryear {%
{Mainzer}%
\ \protect \BOthers {.}}{%
{Mainzer}%
\ \protect \BOthers {.}}{%
{\protect \APACyear {2011}}%
}]{%
2011ApJ...731...53M}
\APACinsertmetastar {%
2011ApJ...731...53M}%
\begin{APACrefauthors}%
{Mainzer}, A.%
, {Bauer}, J.%
, {Grav}, T.%
\ et al.\end{APACrefauthors}%
\unskip\
\newblock
\APACrefYearMonthDay{2011}{{\APACmonth{04}}}{},
\newblock
\unskip
\newblock
\APACjournalVolNumPages{\apj}{731}{1}{53}.
\newblock
\begin{APACrefDOI} \doi{10.1088/0004-637X/731/1/53} \end{APACrefDOI}
\PrintBackRefs{\CurrentBib}

\bibitem [\protect \citeauthoryear {%
{Marshall}%
, {Robin}%
, {Reyl{\'e}}%
, {Schultheis}%
\BCBL {}\ \BBA {} {Picaud}%
}{%
{Marshall}%
\ \protect \BOthers {.}}{%
{\protect \APACyear {2006}}%
}]{%
2006A&A...453..635M}
\APACinsertmetastar {%
2006A&A...453..635M}%
\begin{APACrefauthors}%
{Marshall}, D\BPBI J.%
, {Robin}, A\BPBI C.%
, {Reyl{\'e}}, C.%
, {Schultheis}, M.%
\BCBL {}\ \BBA {} {Picaud}, S.%
\end{APACrefauthors}%
\unskip\
\newblock
\APACrefYearMonthDay{2006}{{\APACmonth{07}}}{},
\newblock
\unskip
\newblock
\APACjournalVolNumPages{\aap}{453}{2}{635-651}.
\newblock
\begin{APACrefDOI} \doi{10.1051/0004-6361:20053842} \end{APACrefDOI}
\PrintBackRefs{\CurrentBib}

\bibitem [\protect \citeauthoryear {%
{Merc}%
}{%
{Merc}%
}{%
{\protect \APACyear {2022}}%
}]{%
2022PhDT........34M}
\APACinsertmetastar {%
2022PhDT........34M}%
\begin{APACrefauthors}%
{Merc}, J.%
\end{APACrefauthors}%
\unskip\
\newblock
\APACrefYear{2022}.
\unskip\
\newblock
\APACrefbtitle {{Multi-frequency research of symbiotic binaries}} {{Multi-frequency research of symbiotic binaries}}\ \APACtypeAddressSchool {PhD Thesis}{}{}.
\unskip\
\newblock
\APACaddressSchool {}{Charles University in Prague / P. J. {\v{S}}af{\'a}rik University in Ko{\v{s}}ice, Czech Republic / Slovakia}.
\PrintBackRefs{\CurrentBib}

\bibitem [\protect \citeauthoryear {%
{Merc}%
, {Barker}%
\BCBL {}\ \BBA {} {G{\'a}lis}%
}{%
{Merc}%
, {Barker}%
\BCBL {}\ \BBA {} {G{\'a}lis}%
}{%
{\protect \APACyear {2022}}%
}]{%
2022RNAAS...6..253M}
\APACinsertmetastar {%
2022RNAAS...6..253M}%
\begin{APACrefauthors}%
{Merc}, J.%
, {Barker}, H.%
\BCBL {}\ \BBA {} {G{\'a}lis}, R.%
\end{APACrefauthors}%
\unskip\
\newblock
\APACrefYearMonthDay{2022}{{\APACmonth{12}}}{},
\newblock
\unskip
\newblock
\APACjournalVolNumPages{Research Notes of the American Astronomical Society}{6}{12}{253}.
\newblock
\begin{APACrefDOI} \doi{10.3847/2515-5172/aca6ee} \end{APACrefDOI}
\PrintBackRefs{\CurrentBib}

\bibitem [\protect \citeauthoryear {%
{Merc}%
, {G{\'a}lis}%
\BCBL {}\ \BBA {} {Wolf}%
}{%
{Merc}%
\ \protect \BOthers {.}}{%
{\protect \APACyear {2019}}%
{\protect \APACexlab {{\protect \BCnt {1}}}}}]{%
2019RNAAS...3...28M}
\APACinsertmetastar {%
2019RNAAS...3...28M}%
\begin{APACrefauthors}%
{Merc}, J.%
, {G{\'a}lis}, R.%
\BCBL {}\ \BBA {} {Wolf}, M.%
\end{APACrefauthors}%
\unskip\
\newblock
\APACrefYearMonthDay{2019{\protect \BCnt {1}}}{{\APACmonth{02}}}{},
\newblock
\unskip
\newblock
\APACjournalVolNumPages{Research Notes of the American Astronomical Society}{3}{2}{28}.
\newblock
\begin{APACrefDOI} \doi{10.3847/2515-5172/ab0429} \end{APACrefDOI}
\PrintBackRefs{\CurrentBib}

\bibitem [\protect \citeauthoryear {%
{Merc}%
, {G{\'a}lis}%
\BCBL {}\ \BBA {} {Wolf}%
}{%
{Merc}%
\ \protect \BOthers {.}}{%
{\protect \APACyear {2019}}%
{\protect \APACexlab {{\protect \BCnt {2}}}}}]{%
2019AN....340..598M}
\APACinsertmetastar {%
2019AN....340..598M}%
\begin{APACrefauthors}%
{Merc}, J.%
, {G{\'a}lis}, R.%
\BCBL {}\ \BBA {} {Wolf}, M.%
\end{APACrefauthors}%
\unskip\
\newblock
\APACrefYearMonthDay{2019{\protect \BCnt {2}}}{{\APACmonth{08}}}{},
\newblock
\unskip
\newblock
\APACjournalVolNumPages{Astronomische Nachrichten}{340}{7}{598-606}.
\newblock
\begin{APACrefDOI} \doi{10.1002/asna.201913662} \end{APACrefDOI}
\PrintBackRefs{\CurrentBib}

\bibitem [\protect \citeauthoryear {%
{Merc}%
\ \protect \BOthers {.}}{%
{Merc}%
\ \protect \BOthers {.}}{%
{\protect \APACyear {2020}}%
}]{%
2020A&A...644A..49M}
\APACinsertmetastar {%
2020A&A...644A..49M}%
\begin{APACrefauthors}%
{Merc}, J.%
, {Miko{\l}ajewska}, J.%
, {Gromadzki}, M.%
\ et al.\end{APACrefauthors}%
\unskip\
\newblock
\APACrefYearMonthDay{2020}{{\APACmonth{12}}}{},
\newblock
\unskip
\newblock
\APACjournalVolNumPages{\aap}{644}{}{A49}.
\newblock
\begin{APACrefDOI} \doi{10.1051/0004-6361/202039132} \end{APACrefDOI}
\PrintBackRefs{\CurrentBib}

\bibitem [\protect \citeauthoryear {%
{Merc}%
\ \protect \BOthers {.}}{%
{Merc}%
\ \protect \BOthers {.}}{%
{\protect \APACyear {2023}}%
}]{%
2023ATel16257....1M}
\APACinsertmetastar {%
2023ATel16257....1M}%
\begin{APACrefauthors}%
{Merc}, J.%
, {Velez}, P.%
, {Barker}, H.%
\ et al.\end{APACrefauthors}%
\unskip\
\newblock
\APACrefYearMonthDay{2023}{{\APACmonth{09}}}{},
\newblock
\unskip
\newblock
\APACjournalVolNumPages{The Astronomer's Telegram}{16257}{}{1}.
\PrintBackRefs{\CurrentBib}

\bibitem [\protect \citeauthoryear {%
{Merc}%
, {Velez}%
\BCBL {}\ \protect \BOthers {.}}{%
{Merc}%
, {Velez}%
\BCBL {}\ \protect \BOthers {.}}{%
{\protect \APACyear {2022}}%
}]{%
2022ATel15340....1M}
\APACinsertmetastar {%
2022ATel15340....1M}%
\begin{APACrefauthors}%
{Merc}, J.%
, {Velez}, P.%
, {Barker}, H.%
\ et al.\end{APACrefauthors}%
\unskip\
\newblock
\APACrefYearMonthDay{2022}{{\APACmonth{04}}}{},
\newblock
\unskip
\newblock
\APACjournalVolNumPages{The Astronomer's Telegram}{15340}{}{1}.
\PrintBackRefs{\CurrentBib}

\bibitem [\protect \citeauthoryear {%
{Miko{\l}ajewska}%
}{%
{Miko{\l}ajewska}%
}{%
{\protect \APACyear {2012}}%
}]{%
2012BaltA..21....5M}
\APACinsertmetastar {%
2012BaltA..21....5M}%
\begin{APACrefauthors}%
{Miko{\l}ajewska}, J.%
\end{APACrefauthors}%
\unskip\
\newblock
\APACrefYearMonthDay{2012}{{\APACmonth{01}}}{},
\newblock
\unskip
\newblock
\APACjournalVolNumPages{Baltic Astronomy}{21}{}{5-12}.
\newblock
\begin{APACrefDOI} \doi{10.1515/astro-2017-0352} \end{APACrefDOI}
\PrintBackRefs{\CurrentBib}

\bibitem [\protect \citeauthoryear {%
{Munari}%
}{%
{Munari}%
}{%
{\protect \APACyear {2019}}%
}]{%
2019arXiv190901389M}
\APACinsertmetastar {%
2019arXiv190901389M}%
\begin{APACrefauthors}%
{Munari}, U.%
\end{APACrefauthors}%
\unskip\
\newblock
\APACrefYearMonthDay{2019}{{\APACmonth{09}}}{},
\newblock
\unskip
\newblock
\APACjournalVolNumPages{arXiv e-prints}{}{}{arXiv:1909.01389}.
\newblock
\begin{APACrefDOI} \doi{10.48550/arXiv.1909.01389} \end{APACrefDOI}
\PrintBackRefs{\CurrentBib}

\bibitem [\protect \citeauthoryear {%
{Murset}%
\ \BBA {} {Nussbaumer}%
}{%
{Murset}%
\ \BBA {} {Nussbaumer}%
}{%
{\protect \APACyear {1994}}%
}]{%
1994A&A...282..586M}
\APACinsertmetastar {%
1994A&A...282..586M}%
\begin{APACrefauthors}%
{Murset}, U.%
\BCBT {}\ \BBA {} {Nussbaumer}, H.%
\end{APACrefauthors}%
\unskip\
\newblock
\APACrefYearMonthDay{1994}{{\APACmonth{02}}}{},
\newblock
\unskip
\newblock
\APACjournalVolNumPages{\aap}{282}{}{586-604}.
\PrintBackRefs{\CurrentBib}

\bibitem [\protect \citeauthoryear {%
{M{\"u}rset}%
\ \BBA {} {Schmid}%
}{%
{M{\"u}rset}%
\ \BBA {} {Schmid}%
}{%
{\protect \APACyear {1999}}%
}]{%
1999A&AS..137..473M}
\APACinsertmetastar {%
1999A&AS..137..473M}%
\begin{APACrefauthors}%
{M{\"u}rset}, U.%
\BCBT {}\ \BBA {} {Schmid}, H\BPBI M.%
\end{APACrefauthors}%
\unskip\
\newblock
\APACrefYearMonthDay{1999}{{\APACmonth{06}}}{},
\newblock
\unskip
\newblock
\APACjournalVolNumPages{\aaps}{137}{}{473-493}.
\newblock
\begin{APACrefDOI} \doi{10.1051/aas:1999105} \end{APACrefDOI}
\PrintBackRefs{\CurrentBib}

\bibitem [\protect \citeauthoryear {%
{Nataf}%
\ \protect \BOthers {.}}{%
{Nataf}%
\ \protect \BOthers {.}}{%
{\protect \APACyear {2013}}%
}]{%
2013ApJ...769...88N}
\APACinsertmetastar {%
2013ApJ...769...88N}%
\begin{APACrefauthors}%
{Nataf}, D\BPBI M.%
, {Gould}, A.%
, {Fouqu{\'e}}, P.%
\ et al.\end{APACrefauthors}%
\unskip\
\newblock
\APACrefYearMonthDay{2013}{{\APACmonth{06}}}{},
\newblock
\unskip
\newblock
\APACjournalVolNumPages{\apj}{769}{2}{88}.
\newblock
\begin{APACrefDOI} \doi{10.1088/0004-637X/769/2/88} \end{APACrefDOI}
\PrintBackRefs{\CurrentBib}

\bibitem [\protect \citeauthoryear {%
{Paunzen}%
\ \BBA {} {Vanmunster}%
}{%
{Paunzen}%
\ \BBA {} {Vanmunster}%
}{%
{\protect \APACyear {2016}}%
}]{%
2016AN....337..239P}
\APACinsertmetastar {%
2016AN....337..239P}%
\begin{APACrefauthors}%
{Paunzen}, E.%
\BCBT {}\ \BBA {} {Vanmunster}, T.%
\end{APACrefauthors}%
\unskip\
\newblock
\APACrefYearMonthDay{2016}{{\APACmonth{03}}}{},
\newblock
\unskip
\newblock
\APACjournalVolNumPages{Astronomische Nachrichten}{337}{3}{239}.
\newblock
\begin{APACrefDOI} \doi{10.1002/asna.201512254} \end{APACrefDOI}
\PrintBackRefs{\CurrentBib}

\bibitem [\protect \citeauthoryear {%
{Pojmanski}%
}{%
{Pojmanski}%
}{%
{\protect \APACyear {1997}}%
}]{%
1997AcA....47..467P}
\APACinsertmetastar {%
1997AcA....47..467P}%
\begin{APACrefauthors}%
{Pojmanski}, G.%
\end{APACrefauthors}%
\unskip\
\newblock
\APACrefYearMonthDay{1997}{{\APACmonth{10}}}{},
\newblock
\unskip
\newblock
\APACjournalVolNumPages{\actaa}{47}{}{467-481}.
\newblock
\begin{APACrefDOI} \doi{10.48550/arXiv.astro-ph/9712146} \end{APACrefDOI}
\PrintBackRefs{\CurrentBib}

\bibitem [\protect \citeauthoryear {%
{Samus'}%
, {Kazarovets}%
, {Durlevich}%
, {Kireeva}%
\BCBL {}\ \BBA {} {Pastukhova}%
}{%
{Samus'}%
\ \protect \BOthers {.}}{%
{\protect \APACyear {2017}}%
}]{%
2017ARep...61...80S}
\APACinsertmetastar {%
2017ARep...61...80S}%
\begin{APACrefauthors}%
{Samus'}, N\BPBI N.%
, {Kazarovets}, E\BPBI V.%
, {Durlevich}, O\BPBI V.%
, {Kireeva}, N\BPBI N.%
\BCBL {}\ \BBA {} {Pastukhova}, E\BPBI N.%
\end{APACrefauthors}%
\unskip\
\newblock
\APACrefYearMonthDay{2017}{{\APACmonth{01}}}{},
\newblock
\unskip
\newblock
\APACjournalVolNumPages{Astronomy Reports}{61}{1}{80-88}.
\newblock
\begin{APACrefDOI} \doi{10.1134/S1063772917010085} \end{APACrefDOI}
\PrintBackRefs{\CurrentBib}

\bibitem [\protect \citeauthoryear {%
{Scargle}%
}{%
{Scargle}%
}{%
{\protect \APACyear {1982}}%
}]{%
1982ApJ...263..835S}
\APACinsertmetastar {%
1982ApJ...263..835S}%
\begin{APACrefauthors}%
{Scargle}, J\BPBI D.%
\end{APACrefauthors}%
\unskip\
\newblock
\APACrefYearMonthDay{1982}{{\APACmonth{12}}}{},
\newblock
\unskip
\newblock
\APACjournalVolNumPages{\apj}{263}{}{835-853}.
\newblock
\begin{APACrefDOI} \doi{10.1086/160554} \end{APACrefDOI}
\PrintBackRefs{\CurrentBib}

\bibitem [\protect \citeauthoryear {%
{Schlafly}%
\ \BBA {} {Finkbeiner}%
}{%
{Schlafly}%
\ \BBA {} {Finkbeiner}%
}{%
{\protect \APACyear {2011}}%
}]{%
2011ApJ...737..103S}
\APACinsertmetastar {%
2011ApJ...737..103S}%
\begin{APACrefauthors}%
{Schlafly}, E\BPBI F.%
\BCBT {}\ \BBA {} {Finkbeiner}, D\BPBI P.%
\end{APACrefauthors}%
\unskip\
\newblock
\APACrefYearMonthDay{2011}{{\APACmonth{08}}}{},
\newblock
\unskip
\newblock
\APACjournalVolNumPages{\apj}{737}{2}{103}.
\newblock
\begin{APACrefDOI} \doi{10.1088/0004-637X/737/2/103} \end{APACrefDOI}
\PrintBackRefs{\CurrentBib}

\bibitem [\protect \citeauthoryear {%
{Seker{\'a}{\v{s}}}%
\ \protect \BOthers {.}}{%
{Seker{\'a}{\v{s}}}%
\ \protect \BOthers {.}}{%
{\protect \APACyear {2019}}%
}]{%
2019CoSka..49...19S}
\APACinsertmetastar {%
2019CoSka..49...19S}%
\begin{APACrefauthors}%
{Seker{\'a}{\v{s}}}, M.%
, {Skopal}, A.%
, {Shugarov}, S.%
\ et al.\end{APACrefauthors}%
\unskip\
\newblock
\APACrefYearMonthDay{2019}{{\APACmonth{04}}}{},
\newblock
\unskip
\newblock
\APACjournalVolNumPages{Contributions of the Astronomical Observatory Skalnate Pleso}{49}{1}{19-66}.
\newblock
\begin{APACrefDOI} \doi{10.48550/arXiv.1904.05555} \end{APACrefDOI}
\PrintBackRefs{\CurrentBib}

\bibitem [\protect \citeauthoryear {%
{Shappee}%
\ \protect \BOthers {.}}{%
{Shappee}%
\ \protect \BOthers {.}}{%
{\protect \APACyear {2014}}%
}]{%
2014ApJ...788...48S}
\APACinsertmetastar {%
2014ApJ...788...48S}%
\begin{APACrefauthors}%
{Shappee}, B\BPBI J.%
, {Prieto}, J\BPBI L.%
, {Grupe}, D.%
\ et al.\end{APACrefauthors}%
\unskip\
\newblock
\APACrefYearMonthDay{2014}{{\APACmonth{06}}}{},
\newblock
\unskip
\newblock
\APACjournalVolNumPages{\apj}{788}{1}{48}.
\newblock
\begin{APACrefDOI} \doi{10.1088/0004-637X/788/1/48} \end{APACrefDOI}
\PrintBackRefs{\CurrentBib}

\bibitem [\protect \citeauthoryear {%
{Shingles}%
\ \protect \BOthers {.}}{%
{Shingles}%
\ \protect \BOthers {.}}{%
{\protect \APACyear {2021}}%
}]{%
2021TNSAN...7....1S}
\APACinsertmetastar {%
2021TNSAN...7....1S}%
\begin{APACrefauthors}%
{Shingles}, L.%
, {Smith}, K\BPBI W.%
, {Young}, D\BPBI R.%
\ et al.\end{APACrefauthors}%
\unskip\
\newblock
\APACrefYearMonthDay{2021}{{\APACmonth{01}}}{},
\newblock
\unskip
\newblock
\APACjournalVolNumPages{Transient Name Server AstroNote}{7}{}{1-7}.
\PrintBackRefs{\CurrentBib}

\bibitem [\protect \citeauthoryear {%
{Smith}%
\ \protect \BOthers {.}}{%
{Smith}%
\ \protect \BOthers {.}}{%
{\protect \APACyear {2020}}%
}]{%
2020PASP..132h5002S}
\APACinsertmetastar {%
2020PASP..132h5002S}%
\begin{APACrefauthors}%
{Smith}, K\BPBI W.%
, {Smartt}, S\BPBI J.%
, {Young}, D\BPBI R.%
\ et al.\end{APACrefauthors}%
\unskip\
\newblock
\APACrefYearMonthDay{2020}{{\APACmonth{08}}}{},
\newblock
\unskip
\newblock
\APACjournalVolNumPages{\pasp}{132}{1014}{085002}.
\newblock
\begin{APACrefDOI} \doi{10.1088/1538-3873/ab936e} \end{APACrefDOI}
\PrintBackRefs{\CurrentBib}

\bibitem [\protect \citeauthoryear {%
{Strai{\v{z}}ys}%
\ \BBA {} {Lazauskait{\.{e}}}%
}{%
{Strai{\v{z}}ys}%
\ \BBA {} {Lazauskait{\.{e}}}%
}{%
{\protect \APACyear {2009}}%
}]{%
2009BaltA..18...19S}
\APACinsertmetastar {%
2009BaltA..18...19S}%
\begin{APACrefauthors}%
{Strai{\v{z}}ys}, V.%
\BCBT {}\ \BBA {} {Lazauskait{\.{e}}}, R.%
\end{APACrefauthors}%
\unskip\
\newblock
\APACrefYearMonthDay{2009}{{\APACmonth{01}}}{},
\newblock
\unskip
\newblock
\APACjournalVolNumPages{Baltic Astronomy}{18}{}{19-31}.
\newblock
\begin{APACrefDOI} \doi{10.48550/arXiv.0907.2398} \end{APACrefDOI}
\PrintBackRefs{\CurrentBib}

\bibitem [\protect \citeauthoryear {%
{Swope}%
\ \BBA {} {Shapley}%
}{%
{Swope}%
\ \BBA {} {Shapley}%
}{%
{\protect \APACyear {1940}}%
}]{%
1940AnHar..90..207S}
\APACinsertmetastar {%
1940AnHar..90..207S}%
\begin{APACrefauthors}%
{Swope}, H\BPBI H.%
\BCBT {}\ \BBA {} {Shapley}, H.%
\end{APACrefauthors}%
\unskip\
\newblock
\APACrefYearMonthDay{1940}{{\APACmonth{01}}}{},
\newblock
\unskip
\newblock
\APACjournalVolNumPages{Annals of Harvard College Observatory}{90}{7}{207-229}.
\PrintBackRefs{\CurrentBib}

\bibitem [\protect \citeauthoryear {%
{Tonry}%
\ \protect \BOthers {.}}{%
{Tonry}%
\ \protect \BOthers {.}}{%
{\protect \APACyear {2018}}%
}]{%
2018PASP..130f4505T}
\APACinsertmetastar {%
2018PASP..130f4505T}%
\begin{APACrefauthors}%
{Tonry}, J\BPBI L.%
, {Denneau}, L.%
, {Heinze}, A\BPBI N.%
\ et al.\end{APACrefauthors}%
\unskip\
\newblock
\APACrefYearMonthDay{2018}{{\APACmonth{06}}}{},
\newblock
\unskip
\newblock
\APACjournalVolNumPages{\pasp}{130}{988}{064505}.
\newblock
\begin{APACrefDOI} \doi{10.1088/1538-3873/aabadf} \end{APACrefDOI}
\PrintBackRefs{\CurrentBib}

\bibitem [\protect \citeauthoryear {%
{Udalski}%
, {Szyma{\'n}ski}%
\BCBL {}\ \BBA {} {Szyma{\'n}ski}%
}{%
{Udalski}%
\ \protect \BOthers {.}}{%
{\protect \APACyear {2015}}%
}]{%
2015AcA....65....1U}
\APACinsertmetastar {%
2015AcA....65....1U}%
\begin{APACrefauthors}%
{Udalski}, A.%
, {Szyma{\'n}ski}, M\BPBI K.%
\BCBL {}\ \BBA {} {Szyma{\'n}ski}, G.%
\end{APACrefauthors}%
\unskip\
\newblock
\APACrefYearMonthDay{2015}{{\APACmonth{03}}}{},
\newblock
\unskip
\newblock
\APACjournalVolNumPages{\actaa}{65}{1}{1-38}.
\newblock
\begin{APACrefDOI} \doi{10.48550/arXiv.1504.05966} \end{APACrefDOI}
\PrintBackRefs{\CurrentBib}

\bibitem [\protect \citeauthoryear {%
{Whitelock}%
}{%
{Whitelock}%
}{%
{\protect \APACyear {2003}}%
}]{%
2003ASPC..303...41W}
\APACinsertmetastar {%
2003ASPC..303...41W}%
\begin{APACrefauthors}%
{Whitelock}, P\BPBI A.%
\end{APACrefauthors}%
\unskip\
\newblock
\APACrefYearMonthDay{2003}{{\APACmonth{01}}}{},
\newblock
{\BBOQ}\APACrefatitle {{A Comparison of Symbiotic and Normal Miras (invited review talks)}} {{A Comparison of Symbiotic and Normal Miras (invited review talks)}}.{\BBCQ}
\newblock
\BIn{} R\BPBI L\BPBI M.~{Corradi}, J.~{Mikolajewska}\BCBL {}\ \BBA {} T\BPBI J.~{Mahoney}\ (\BEDS), \APACrefbtitle {Symbiotic Stars Probing Stellar Evolution} {Symbiotic Stars Probing Stellar Evolution}\ \BVOL~303, \BPG~41.
\PrintBackRefs{\CurrentBib}

\bibitem [\protect \citeauthoryear {%
{Wright}%
\ \protect \BOthers {.}}{%
{Wright}%
\ \protect \BOthers {.}}{%
{\protect \APACyear {2010}}%
}]{%
2010AJ....140.1868W}
\APACinsertmetastar {%
2010AJ....140.1868W}%
\begin{APACrefauthors}%
{Wright}, E\BPBI L.%
, {Eisenhardt}, P\BPBI R\BPBI M.%
, {Mainzer}, A\BPBI K.%
\ et al.\end{APACrefauthors}%
\unskip\
\newblock
\APACrefYearMonthDay{2010}{{\APACmonth{12}}}{},
\newblock
\unskip
\newblock
\APACjournalVolNumPages{\aj}{140}{6}{1868-1881}.
\newblock
\begin{APACrefDOI} \doi{10.1088/0004-6256/140/6/1868} \end{APACrefDOI}
\PrintBackRefs{\CurrentBib}

\end{thebibliography}



\end{document}